\def\arxiv{}
\documentclass[conference,9pt]{IEEEtran}  %

\usepackage{ulem}
\usepackage{iftex}
\ifPDFTeX
  \usepackage{courier}                                     %
\else
  \usepackage{fontspec}
\fi
\usepackage{textcomp}                                      %
\usepackage{graphicx}
\usepackage{amsmath}
\usepackage{amssymb}
\usepackage{amsthm}                                        %
\usepackage{mathtools}
\usepackage{comment}
\usepackage[caption=false,font=footnotesize]{subfig}
\usepackage{bm}
\usepackage{multirow}
\usepackage{threeparttable,booktabs}
\usepackage{tikz}
\usepackage{balance}
\usepackage[mathcal]{eucal}
\usepackage[noend]{algpseudocode}
\algrenewcommand\textproc{\texttt}
\usepackage{algorithm}
\usepackage{siunitx}
\usepackage{cite}                                          %

\theoremstyle{plain}

\theoremstyle{definition}

 \graphicspath{{./figs/}}

\usepackage{color}
\usepackage{xcolor}
\usepackage{colortbl}%
\definecolor{failcell}{gray}{0.88}%
\definecolor{ourcell}{RGB}{236,242,252}%
\definecolor{groupcell}{gray}{0.93}%
\usepackage{makecell}
\usepackage{amsmath}
\usepackage{graphicx}
\usepackage{hyperref}
\hypersetup{pdfauthor={},pdftitle={},pdfsubject={},pdfkeywords={}}
\usepackage{cleveref}                                      %
\usepackage{booktabs}
\usepackage{multirow}
\usepackage{threeparttable}
\usepackage{rotating}
\newcommand{\sys}{CleanRoute}   %

\usetikzlibrary{arrows.meta,positioning,calc,fit,patterns}
\newcommand{\CapFigInfo}{Three key problems of analog routing, each with a layout that fails it (top) and the layout of \sys{} (bottom).}
\newcommand{\CapFigFlow}{\sys{} flow, with the steps of its three techniques numbered.}
\newcommand{\CapFigMech}{A DRC violation of each rule class (top) and the same layout window in \sys{} (bottom).}%
\newcommand{\CapFigGrid}{Uniform grids of step $2S$ (a) and $S$ (b) leave DRC violations, while the $2S$ grid plus an edge-aligned grid (c) leaves none, with runtimes relative to (c).}
\newcommand{\CapFigTree}{Supply wires of a current mirror (reference Ref, copy Mir) and of a high-current device outside it, shared (a) and as a supply tree (b).}%
\newcommand{\CapTabCoverage}{Rule families modeled by each router.}
\newcommand{\NoteTabCoverage}{\fullmark{} modeled in full, \halfmark{} in part, \nomark{} not at all, \nodesc{} unknown. Rows are grouped by class, the stage at which \sys{} checks a family. PRL, parallel run length. EOL, end-of-line.}

\newcommand{\CapTabBOne}{DRC, LVS, runtime and performance comparison with existing analog routers.}
\newcommand{\NoteTabBOne}{Gray means no simulatable layout. A dash means the router fails to complete the routing ($^{a}$) or the layout is LVS-incorrect ($^{e}$). Bold means the better post-layout value.\SetMarkNote{} Anaroute routes MAGICAL's PCells at the device centers of APPlace.\csname NoteBSum@one\endcsname}
\expandafter\newcommand\csname NoteBSum@one\endcsname{}

\newcommand{\CapTabAblThree}{Ablation study of the three techniques.}
\newcommand{\NoteTabAblThree}{Shorts are those of the final check. Sums over the ten circuits, except LVS, Stable, and PM, which count circuits, and runtime and offset, which are medians, the offset over the functional layouts after extraction.}

\newcommand{\CapTabSwapS}{Ablation study on MAGICAL's placements.}%
\newcommand{\NoteTabSwap}{Gray means no simulatable layout.\SwapDashNote{}\SwapBoldNote{}\SetMarkNote{}}%

\expandafter\newcommand\csname CapFigRFive@c\endcsname{Criteria met by each router, in circuits of ten.}
\expandafter\newcommand\csname CapFigStats@c2\endcsname{Overall outcomes of the three routers on the 11 circuits in TSMC 40\,nm.}

\newcommand{\CapTabComp}{Comparison with existing analog routers on a StrongARM comparator.}
\newcommand{\NoteTabComp}{Bold means the best value among the routers.}
\newcommand{\CapTabSky}{Generalization of \sys{} to \OthSkyN{} circuits in SkyWater 130\,nm.}
\expandafter\newcommand\csname NoteTabSky@pc\endcsname{Runtime is the median routing runtime of three runs.}
\expandafter\newcommand\csname NoteTabSky@bt\endcsname{DRC-clean means no violations reported by Magic or KLayout. Runtimes are the medians of three runs per circuit.}
\expandafter\newcommand\csname NoteTabOther@full\endcsname{Bold means the best value among the routers.}
\newcommand{\CapFigIrdrop}{Largest source-voltage difference between a current-mirror reference and its copies on the ground net, from the post-layout DC operating point.}
\DeclareRobustCommand{\fullmark}{\checkmark}
\DeclareRobustCommand{\halfmark}{\tikz[baseline=-0.6ex]{\fill (0,0) -- (90:0.6ex) arc[start angle=90,end angle=270,radius=0.6ex] -- cycle;\draw[line width=0.4pt] (0,0) circle[radius=0.6ex];}}
\DeclareRobustCommand{\nomark}{$\times$}
\DeclareRobustCommand{\nodesc}{--}

\newcommand{\SetMark}{}
\newcommand{\IfDag}[1]{}%
\newcommand{\SetMarkNote}{}
\newsavebox{\wnTabBox}
\newcommand{\wnTab}[2]{\sbox{\wnTabBox}{#2}\typeout{TABWD #1 = \the\wd\wnTabBox}\usebox{\wnTabBox}}
\newcommand{\TabNoteFont}{\fontsize{6}{7}\selectfont}
\newcommand{\wnTabN}[3]{\sbox{\wnTabBox}{#2}\typeout{TABWD #1 = \the\wd\wnTabBox}\begin{minipage}{\wd\wnTabBox}\usebox{\wnTabBox}\par\vspace{1.5pt}{\TabNoteFont\raggedright #3\par}\end{minipage}\par}
\newcommand{\TabCoverage}{%
\begin{table}[tb]
\centering
\caption{\CapTabCoverage}
\label{tab:coverage}
\fontsize{7}{8.6}\selectfont%
\setlength{\tabcolsep}{0.8pt}
\wnTabN{cov}{\begin{tabular}{@{}lccccc@{}}
\toprule
Rule family & Anaroute~\cite{chen_magicalroute_2020} & SAGERoute~\cite{zhang_sageroute_2023} & ALIGN~\cite{kunal_align_2019} & GeoRoute~\cite{pan_georoute_2026} & \sys{} \\
\midrule
\rowcolor{groupcell}\multicolumn{6}{@{}l}{1) \emph{Step-level}} \\
PRL spacing & \halfmark & \halfmark & \halfmark & \halfmark & \fullmark \\
Same-net spacing & \halfmark & \nomark & \nomark & \halfmark & \fullmark \\
Via enclosure & \nomark & \nomark & \halfmark & \nodesc & \fullmark \\
Via spacing, count & \halfmark & \nomark & \halfmark & \halfmark & \fullmark \\
\rowcolor{groupcell}\multicolumn{6}{@{}l}{2) \emph{Net-level}} \\
Min. area & \fullmark & \nomark & \nomark & \nodesc & \fullmark \\
\rowcolor{groupcell}\multicolumn{6}{@{}l}{3) \emph{Layout-level}} \\
EOL spacing & \nomark & \nomark & \halfmark & \nodesc & \fullmark \\
Short edge & \nomark & \nomark & \nomark & \nodesc & \fullmark \\
Min. width & \fullmark & \fullmark & \fullmark & \nodesc & \fullmark \\
\rowcolor{groupcell}\multicolumn{6}{@{}l}{4) \emph{Device-level}} \\
Mfg. grid & \fullmark & \nomark & \nomark & \nodesc & \fullmark \\
Device keep-out & \fullmark & \nomark & \halfmark & \halfmark & \fullmark \\
\midrule
Met in full & 4/10 & 1/10 & 1/10 & 0/10 & 10/10 \\
\bottomrule
\end{tabular}}{\NoteTabCoverage}%
\vspace{-0.3cm}
\end{table}}

\newcommand{\BOneFont}{\fontsize{7}{8.2}\selectfont}
\newcommand{\TabBOneOne}{%
\begin{table*}[tb]
\centering
\caption{\CapTabBOne}
\label{tab:res}\label{tab:post}
\BOneFont
\setlength{\tabcolsep}{1.6pt}
\renewcommand{\arraystretch}{0.92}
\wnTabN{b1_one}{%
\begin{tabular}{@{}l@{\hskip 7pt}rrr@{\hskip 7pt}rcrrrrrr@{\hskip 7pt}rcr@{\hskip 7pt}rcrrrrrr@{}}
\toprule
 & \multicolumn{3}{c}{Schematic} & \multicolumn{8}{c}{Anaroute~\cite{chen_magicalroute_2020}} & \multicolumn{3}{c}{SAGERoute~\cite{zhang_sageroute_2023}} & \multicolumn{8}{c}{\sys{}} \\
\cmidrule(lr){2-4}\cmidrule(lr){5-12}\cmidrule(lr){13-15}\cmidrule(l){16-23}
Circuit & PM & UGB & Gain & DRC & LVS & Runtime & PM & UGB & Gain & CMRR & Offset & DRC & LVS & Runtime & DRC & LVS & Runtime & PM & UGB & Gain & CMRR & Offset \\
 & (deg) & (MHz) & (dB) & viol. &  & (s) & (deg) & (MHz) & (dB) & (dB) & (\textmu V) & viol. &  & (s) & viol. &  & (s) & (deg) & (MHz) & (dB) & (dB) & (\textmu V) \\
\midrule
ota1 & 67.3 & 2.11 & 49.6 & 270 & \checkmark & 5.22 & 67.1 & \textbf{1.98} & 49.4 & \textbf{84.8} & \textbf{1} & \cellcolor{failcell}2757 & \cellcolor{failcell}$\times$ & \cellcolor{failcell}56.5 & 0 & \checkmark & 1.45 & \textbf{67.3} & 1.94 & \textbf{49.5} & 83.9 & 64 \\
ota2 & 68.8 & 5.47 & 62.0 & \cellcolor{failcell}--\smash{$^{a}$} & \cellcolor{failcell}--\smash{$^{a}$} & \cellcolor{failcell}--\smash{$^{a}$} & \cellcolor{failcell}--\smash{$^{a}$} & \cellcolor{failcell}--\smash{$^{a}$} & \cellcolor{failcell}--\smash{$^{a}$} & \cellcolor{failcell}--\smash{$^{a}$} & \cellcolor{failcell}--\smash{$^{a}$} & \cellcolor{failcell}6040 & \cellcolor{failcell}$\times$ & \cellcolor{failcell}120.3 & 0 & \checkmark & 11.3 & \textbf{65.1} & \textbf{5.13} & \textbf{63.4} & \textbf{68.8} & \textbf{47} \\
ota3 & 68.7 & 11.09 & 65.7 & \cellcolor{failcell}654 & \cellcolor{failcell}$\times$ & \cellcolor{failcell}29.0 & \cellcolor{failcell}--\smash{$^{e}$} & \cellcolor{failcell}--\smash{$^{e}$} & \cellcolor{failcell}--\smash{$^{e}$} & \cellcolor{failcell}--\smash{$^{e}$} & \cellcolor{failcell}--\smash{$^{e}$} & \cellcolor{failcell}5656 & \cellcolor{failcell}$\times$ & \cellcolor{failcell}160.4 & 0 & \checkmark & 5.85 & \textbf{63.0} & \textbf{10.63} & \textbf{68.3} & \textbf{81.4} & \textbf{44} \\
ota4 & 61.3 & 12.70 & 65.5 & \cellcolor{failcell}860 & \cellcolor{failcell}$\times$ & \cellcolor{failcell}38.6 & \cellcolor{failcell}--\smash{$^{e}$} & \cellcolor{failcell}--\smash{$^{e}$} & \cellcolor{failcell}--\smash{$^{e}$} & \cellcolor{failcell}--\smash{$^{e}$} & \cellcolor{failcell}--\smash{$^{e}$} & \cellcolor{failcell}5514 & \cellcolor{failcell}$\times$ & \cellcolor{failcell}148.6 & 0 & \checkmark & 5.13 & \textbf{50.3} & \textbf{11.53} & \textbf{68.0} & \textbf{71.5} & \textbf{184} \\
ota5 & 70.7 & 10.47 & 64.4 & \cellcolor{failcell}983 & \cellcolor{failcell}$\times$ & \cellcolor{failcell}87.6 & \cellcolor{failcell}--\smash{$^{e}$} & \cellcolor{failcell}--\smash{$^{e}$} & \cellcolor{failcell}--\smash{$^{e}$} & \cellcolor{failcell}--\smash{$^{e}$} & \cellcolor{failcell}--\smash{$^{e}$} & \cellcolor{failcell}5588 & \cellcolor{failcell}$\times$ & \cellcolor{failcell}85.0 & 0 & \checkmark & 10.3 & \textbf{70.9} & \textbf{7.79} & \textbf{67.5} & \textbf{67.4} & \textbf{17} \\
ota6 & 69.3 & 11.49 & 73.5 & \cellcolor{failcell}753 & \cellcolor{failcell}$\times$ & \cellcolor{failcell}3.97 & \cellcolor{failcell}--\smash{$^{e}$} & \cellcolor{failcell}--\smash{$^{e}$} & \cellcolor{failcell}--\smash{$^{e}$} & \cellcolor{failcell}--\smash{$^{e}$} & \cellcolor{failcell}--\smash{$^{e}$} & \cellcolor{failcell}--\smash{$^{a}$} & \cellcolor{failcell}--\smash{$^{a}$} & \cellcolor{failcell}--\smash{$^{a}$} & 0 & \checkmark & 7.29 & \textbf{60.5} & \textbf{7.71} & \textbf{74.0} & \textbf{42.7} & \textbf{407} \\
ota7 & 70.7 & 15.19 & 82.7 & \cellcolor{failcell}865\SetMark{} & \cellcolor{failcell}$\times$ & \cellcolor{failcell}51.9 & \cellcolor{failcell}--\smash{$^{e}$} & \cellcolor{failcell}--\smash{$^{e}$} & \cellcolor{failcell}--\smash{$^{e}$} & \cellcolor{failcell}--\smash{$^{e}$} & \cellcolor{failcell}--\smash{$^{e}$} & \cellcolor{failcell}--\smash{$^{a}$} & \cellcolor{failcell}--\smash{$^{a}$} & \cellcolor{failcell}--\smash{$^{a}$} & 0 & \checkmark & 5.20 & \textbf{62.4} & \textbf{10.50} & \textbf{80.8} & \textbf{81.2} & \textbf{4} \\
ota8 & 74.7 & 17.43 & 81.8 & 814\SetMark{} & \checkmark & 19.4 & 64.2 & 10.35 & 76.8 & \textbf{90.3} & \textbf{1} & \cellcolor{failcell}--\smash{$^{a}$} & \cellcolor{failcell}--\smash{$^{a}$} & \cellcolor{failcell}--\smash{$^{a}$} & 0 & \checkmark & 23.8 & \textbf{69.0} & \textbf{10.58} & \textbf{80.2} & 54.1 & 55 \\
ota9 & 65.6 & 17.80 & 80.6 & \cellcolor{failcell}1394\SetMark{} & \cellcolor{failcell}$\times$ & \cellcolor{failcell}5.34 & \cellcolor{failcell}--\smash{$^{e}$} & \cellcolor{failcell}--\smash{$^{e}$} & \cellcolor{failcell}--\smash{$^{e}$} & \cellcolor{failcell}--\smash{$^{e}$} & \cellcolor{failcell}--\smash{$^{e}$} & \cellcolor{failcell}--\smash{$^{a}$} & \cellcolor{failcell}--\smash{$^{a}$} & \cellcolor{failcell}--\smash{$^{a}$} & 0 & \checkmark & 11.2 & \textbf{54.4} & \textbf{10.73} & \textbf{77.3} & \textbf{98.5} & \textbf{10} \\
ota10 & 70.6 & 14.69 & 83.0 & \cellcolor{failcell}1196\SetMark{} & \cellcolor{failcell}$\times$ & \cellcolor{failcell}7.28 & \cellcolor{failcell}--\smash{$^{e}$} & \cellcolor{failcell}--\smash{$^{e}$} & \cellcolor{failcell}--\smash{$^{e}$} & \cellcolor{failcell}--\smash{$^{e}$} & \cellcolor{failcell}--\smash{$^{e}$} & \cellcolor{failcell}--\smash{$^{a}$} & \cellcolor{failcell}--\smash{$^{a}$} & \cellcolor{failcell}--\smash{$^{a}$} & 0 & \checkmark & 15.2 & \textbf{60.1} & \textbf{10.22} & \textbf{80.4} & \textbf{47.6} & \textbf{295} \\
\bottomrule
\end{tabular}
}{\NoteTabBOne}%
\vspace{-0.3cm}
\end{table*}}

\newcommand{\TabBOne}{\csname TabBOne@one\endcsname}
\expandafter\let\csname TabBOne@one\endcsname\TabBOneOne

\newcommand{\TabAblThree}{%
\begin{table}[tb]
\centering
\caption{\CapTabAblThree}
\label{tab:abl}
\fontsize{7}{8.2}\selectfont
\setlength{\tabcolsep}{1.2pt}%
\renewcommand{\arraystretch}{0.92}
\wnTabN{abl3g}{%
\begin{tabular}{@{}llrrcr@{}}
\toprule
Stage & Mechanism off & DRC viol. & Shorts & LVS & Runtime (s) \\
\midrule
\multicolumn{2}{@{}l}{\sys{} (full)} & 0 & 0 & 10 & 8.82 \\
\midrule
\rowcolor{groupcell}\multicolumn{6}{@{}l}{1) \textit{Hierarchical rule enforcement}} \\
Step-level & partial deferral & 1 & 5 & 9 & 10.1 \\
 & via escape & 12 & 0 & 10 & 33.1 \\
 & via arrays & 20 & 0 & 10 & 16.9 \\
Net-level & pad expansion & 228 & 0 & 10 & 8.79 \\
Layout-level & short-edge fill & 14 & 0 & 10 & 8.75 \\
 & notch fill & 16 & 0 & 10 & 9.43 \\
 & line-end extension & 2 & 0 & 10 & 7.67 \\
 & wire shifting & 6 & 0 & 10 & 8.04 \\
Device-level & capacitor keep-out & 8 & 0 & 9 & 5.00 \\
\midrule
\rowcolor{groupcell}\multicolumn{6}{@{}l}{2) \textit{Topology-based mismatch mitigation}} \\
 & supply tree & 0 & 0 & 10 & 1.87 \\
 & trunks & 1 & 0 & 10 & 3.90 \\
\cmidrule(l){3-6}
 & \textit{After extraction} & Stable & PM$\,\geq$45 & PM$\,\geq$60 & Offset (\textmu V) \\
 & \sys{} (full) & 10 & 10 & 8 & 51.0 \\
 & supply tree & 8 & 7 & 5 & 129 \\
 & trunks & 10 & 10 & 5 & 44 \\
\midrule
\rowcolor{groupcell}\multicolumn{6}{@{}l}{3) \textit{Hybrid routing grid strategy}} \\
 & edge-aligned grid & 223 & 0 & 7 & 255.0 \\
 & step $S$ for $2S$ & 10 & 0 & 10 & 126.6 \\
 & edge-aligned grid, step $S$ & 341 & 0 & 10 & 455.7 \\
\bottomrule
\end{tabular}
}{\NoteTabAblThree}%
\vspace{-0.3cm}
\end{table}}
\newcommand{\TabSwapBody}[1]{\fontsize{7}{8.2}\selectfont\setlength{\tabcolsep}{#1}\renewcommand{\arraystretch}{0.92}%
\wnTabN{swap_sb}{%
\gdef\SwapMarks{}%
\gdef\SwapDashNote{}%
\gdef\SwapBoldNote{ Bold means the better post-layout value.}%
\gdef\SwapStNote{A router with the only functional layout of a circuit has the better value. Runtime is the median routing runtime. Bold means the better statistic.}%
\begin{tabular}{@{}l@{\hskip 4pt}ccc@{\hskip 4pt}c@{\hskip 5pt}ccccc@{}}
\toprule
 & \multicolumn{3}{c}{Circuits} & & \multicolumn{5}{c}{Circuits with the better value} \\
\cmidrule(lr){2-4}\cmidrule(l){6-10}
Router & \makecell{DRC-\\clean} & \makecell{LVS-\\correct} & Functional & \makecell{Runtime\\(s)} & PM & UGB & Gain & CMRR & Offset \\
\midrule
Anaroute~\cite{chen_magicalroute_2020} & 0/10 & 5/10 & 5/10 & 3.20 & 4/10 & 2/10 & 1/10 & 2/10 & 2/10 \\
\sys{} & \textbf{9/10} & \textbf{10/10} & \textbf{10/10} & 8.91 & \textbf{6/10} & \textbf{8/10} & \textbf{9/10} & \textbf{8/10} & \textbf{8/10} \\
\bottomrule
\end{tabular}
}{\ifdefined\SwapStNote\SwapStNote\else\NoteTabSwap\fi}}%

\newcommand{\TabSwapCol}{%
\begin{table}[tb]
\centering
\caption{\CapSwapSel}%
\label{tab:swap}
\TabSwapBody{1.0pt}
\vspace{-0.3cm}
\end{table}}

\def\CapSwapSel{\CapTabSwapS}

\newcommand{\TabSwap}{\TabSwapCol}

\newcommand{\BlockTwo}{\TabAblThree\TabSwap}

\newcommand{\TabComp}{%
\begin{table}[tb]
\centering
\caption{\CapTabComp}
\label{tab:comp}
\fontsize{7}{8.2}\selectfont\setlength{\tabcolsep}{2.0pt}\renewcommand{\arraystretch}{0.92}%
\wnTabN{comp}{%
\begin{tabular}{@{}lcccccc@{}}
\toprule
Router & DRC viol. & LVS & Runtime (s) & Delay (ps) & Power (\textmu W) & $|$Offset$|$ (mV) \\
\midrule
Schematic & -- & -- & -- & 51.5 & 5.89 & -- \\
Anaroute~\cite{chen_magicalroute_2020} & 126 & \checkmark & \textbf{0.81} & 103.7 & 8.97 & \textbf{2.71} \\
SAGERoute~\cite{zhang_sageroute_2023} & 2241 & \checkmark & 3.82 & 129.6 & 9.19 & 8.91 \\
\sys{} & \textbf{0} & \checkmark & 1.44 & \textbf{98.3} & \textbf{8.91} & 2.95 \\
\bottomrule
\end{tabular}
}{\NoteTabComp}
\vspace{-0.3cm}
\end{table}}
\newcommand{\TabOther}{%
\begin{table}[tb]
\centering
\caption{\CapTabSky}
\label{tab:other}
\fontsize{7}{8.2}\selectfont\setlength{\tabcolsep}{\csname SkySep@pc\endcsname}\renewcommand{\arraystretch}{0.92}%
\begin{tabular}{@{}l@{\hskip 3.5pt}c@{\hskip 3.5pt}c@{\hskip 3.5pt}c@{\hskip 3.5pt}r@{\hskip 5pt}|@{\hskip 5pt}l@{\hskip 3.5pt}c@{\hskip 3.5pt}c@{\hskip 3.5pt}c@{\hskip 3.5pt}r@{}}
\toprule
Circuit & Type & \makecell{DRC\\viol.} & LVS & \makecell{Runtime\\(s)} & Circuit & Type & \makecell{DRC\\viol.} & LVS & \makecell{Runtime\\(s)} \\
\midrule
ota5t & Amp. & 0 & \checkmark & 0.006 & fc1\_fd & Amp. & 0 & \checkmark & 21.2 \\
lb1s\_wsc & Amp. & 0 & \checkmark & 0.021 & amp3\_tcfc & Amp. & 0 & \checkmark & 1.02 \\
fc\_se & Amp. & 0 & \checkmark & 0.91 & amp3\_dacfc & Amp. & 0 & \checkmark & 5.21 \\
nauta & Amp. & 0 & \checkmark & 1.00 & fc2 & Amp. & 0 & \checkmark & 59.9 \\
lb2s\_wsc & Amp. & 0 & \checkmark & 0.74 & otaC & Amp. & 0 & \checkmark & 51.1 \\
lb2s\_pcas & Amp. & 0 & \checkmark & 0.37 & otaB & Amp. & 0 & \checkmark & 29.2 \\
lb1s\_tele & Amp. & 0 & \checkmark & 1.85 & cmp\_ct & Comp. & 0 & \checkmark & 4.84 \\
amp3\_smc & Amp. & 0 & \checkmark & 0.27 & sacmp & Comp. & 0 & \checkmark & 3.27 \\
otaA & Amp. & 0 & \checkmark & 6.95 & ldo & LDO & 0 & \checkmark & 0.56 \\
amp3\_acbc & Amp. & 0 & \checkmark & 4.39 & vco & VCO & 0 & \checkmark & 0.93 \\
\bottomrule
\end{tabular}
\vspace{-0.3cm}
\end{table}}
\expandafter\def\csname SkySep@pc\endcsname{0.8pt}
\expandafter\def\csname SkySep@bt\endcsname{3pt}

\newcommand{\FigStats}{
\begin{figure}[tb]
  \centering
  \includegraphics[width=\columnwidth]{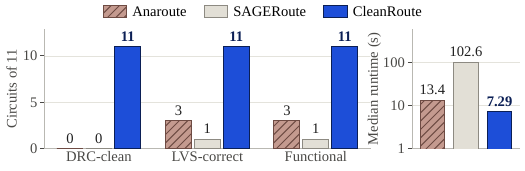}
  \caption{\csname CapFigStats@c2\endcsname}
  \label{fig:stats}
  \vspace{-0.3cm}
\end{figure}}

\newcommand{\AblBaseDRC}{0}

\newcommand{\AblBaseTime}{8.82}

\newcommand{\AblFlushDRC}{223}
\newcommand{\AblFlushLVS}{7}

\newcommand{\AblFlushTime}{255.0}
\newcommand{\AblGridSDRC}{10}

\newcommand{\AblGridSTime}{126.6}

\newcommand{\AblNoFlushSDRC}{341}

\newcommand{\AblNoFlushSTime}{455.7}

\newcommand{\AblPadDRC}{228}

\newcommand{\AblPostBasePMsixty}{8}

\newcommand{\AblPostTreePMtarget}{7}

\newcommand{\AblPostTreePairBaseOffMedPr}{51.0}
\newcommand{\AblPostTreePairN}{8}
\newcommand{\AblPostTreePairOffMed}{129}
\newcommand{\AblPostTreeStable}{8}

\newcommand{\AblPostTrunkPMsixty}{5}

\newcommand{\AblStrictDRC}{1}

\newcommand{\SelAmps}{10}

\newcommand{\SelCircuits}{10}

\newcommand{\SelMPCleanN}{9}
\newcommand{\SelMPDirtyDRC}{7}
\newcommand{\SelMPDirtyName}{ota6}
\newcommand{\SelMPFunc}{10}

\newcommand{\SelRouterLVSDRCMax}{814}
\newcommand{\SelRouterLVSDRCMin}{270}

\newcommand{\SelRouterOnlyUsable}{8}

\newcommand{\SelRouterUsableMagP}{2}

\newcommand{\SelRtFasterMagPDRCMin}{753}

\newcommand{\SelRtFasterMagPN}{4}

\newcommand{\SelRtMagPMed}{19.4}

\newcommand{\SelRtSageMed}{120.3}

\newcommand{\SelRtTreeMed}{8.82}

\newcommand{\SelSageRoutMax}{6040}

\newcommand{\SelSageRoutMin}{2757}

\newcommand{\SelTreeDCchgAbsMax}{3.3}

\newcommand{\IrBaseMed}{0.86}
\newcommand{\IrTreeMed}{26.7}

\newcommand{\IrLowerN}{all ten}
\newcommand{\AllCircN}{31}
\newcommand{\AllRtMed}{4.84}

\newcommand{\OthSkyN}{20}

\newcommand{\OthSkyRtMed}{1.4}
\makeatletter
\let\wnvii@endfloatbox\@endfloatbox
\def\@endfloatbox{\wnvii@endfloatbox\typeout{FLOATHT \@captype\space\csname the\@captype\endcsname\space = \the\ht\@currbox\space+ \the\dp\@currbox}}
\makeatother

\title{\sys{}: Reconciling DRC Compliance, Post-Layout Performance, and Routing Efficiency in Analog Routing}
\ifdefined\arxiv
\author{
  \IEEEauthorblockN{Bingyang Liu, Ziming Wei, Xiaohan Gao, Yang Hsu, and David Z. Pan}
  \IEEEauthorblockA{Department of Electrical and Computer Engineering, The University of Texas at Austin\\
  Email: \{bingyangliu@, weizm@, xiaohan.gao@austin., yanghsu@, dpan@ece.\}utexas.edu}
}
\fi

\begin{document}

\pagestyle{plain}

\sloppy
\ifdefined\arxiv\else
\makeatletter
\def\@IEEEauthorblockconfadjspace{-2.4em}
\def\@IEEENORMtitlevspace{0pt}
\def\@IEEEMINtitlevspace{0pt}
\makeatother
\fi
\maketitle
\begin{abstract}
Automated analog routing remains a long-standing challenge because design-rule checking (DRC) compliance, post-layout performance preservation, and routing efficiency are difficult to achieve together. Existing analog routers struggle to meet all three objectives due to limited design-rule coverage, insufficient control over topology-induced mismatch, or trade-offs between DRC violations and runtime on uniform routing grids.  In this paper, we introduce \sys, an analog router that jointly addresses these challenges while producing DRC-clean layouts. Hierarchical rule enforcement assigns DRC checks and repairs to routing stages, covering rules both within routing metal and where it joins device metal. Topology-based mismatch mitigation constructs supply trees from current-mirror relationships and device currents to reduce topology-induced mismatch that geometric matching cannot eliminate. A hybrid routing grid strategy adds a net-specific grid aligned with device-metal edges to a coarse routing grid, balancing DRC compliance and routing efficiency without extensive manual post-routing repair. Experiments on a comprehensive benchmark of \AllCircN{} circuits of different types in a TSMC 40\,nm process and a 130\,nm process show that \sys{} produces DRC-clean layouts that pass layout versus schematic (LVS) checking on all benchmarks with a median routing runtime of \AllRtMed{}\,s. In particular, in the TSMC 40\,nm process, \sys{} produces functional layouts with the strongest overall performance preservation, whereas the baselines produce no DRC-clean layouts. 
\end{abstract}
\section{Introduction}
\label{sec:intro}

Layout design remains a bottleneck in analog design automation, particularly routing, which targets to balance three competing objectives. First, a routed layout should pass design-rule checking (DRC) to ensure manufacturability and layout versus schematic (LVS) checking, which verifies the layout connectivity against the schematic. Second, routing should preserve the circuit's post-layout performance by limiting the mismatch and other layout-dependent effects introduced by interconnect parasitics and resistance. Third, routing should be fast enough to support repeated iterations of analog layout design for further performance improvement.

\begin{figure}[tb]
  \vspace{-0.2cm}%
  \centering
  \includegraphics[width=\columnwidth]{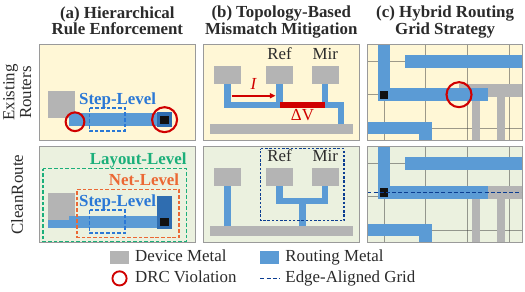}%
  \caption{\CapFigInfo}
  \label{fig:info}
  \vspace{-0.3cm}
\end{figure}

Existing analog routers face distinct difficulties in meeting these objectives. For DRC compliance, a major challenge is that routing should satisfy not only the spacing constraints between separate metal shapes but also the rules governing the geometry formed when metal joins. A layout contains \emph{device metal}, which is fixed before routing and belongs to PCells, guard rings, and power stripes, and \emph{routing metal}, which is added by the router. Their interactions occur wherever routing metal connects to device metal at pins or to other routing metal at branches and vias, and can introduce violations such as short-edge and notch. Existing analog routers~\cite{chen_magicalroute_2020,xu_magical_2019,xu_paroute_2024,xu_paroute2_2025,zhang_sageroute_2023,kunal_align_2019} primarily enforce spacing rules in routing metal, with limited treatment of the DRC violations where routing metal joins device metal. Prior flows~\cite{xu_magical_2019,kunal_align_2019,chang_bag2_2018,han_laygo_2021} alleviate the difficulty at device terminals through integrated device generators that align pins with routing tracks, but the remaining violations at these joints still require attention. Under the PCells and the complex design rules of a commercial PDK, such limitations can lead to DRC violations. As illustrated in Fig.~\ref{fig:info}(a), a wire may pass the check of every search move while violations remain on its finished shape and where it joins device metal. Thus, a more systematic DRC solution is needed to achieve DRC-clean layouts in more realistic scenarios.

Post-layout performance poses a different challenge. Existing routers have extensively explored symmetric routing~\cite{malavasi_automation_1996,malavasi_area_1993,choudhury_channel_1993,chen_magicalroute_2020,zhang_sageroute_2023,kunal_align_2019,wang_mgaroute_2025}, length matching~\cite{ou_matching_2014}, and wire sizing~\cite{ou_current_2013,torabi_ilp_2018,zhang_sageroute_2023} to control geometry-induced mismatch and interconnect resistance. Learning-based methods further guide routing with knowledge from manual layouts, predicted post-layout performance, or estimated parasitic capacitance~\cite{zhu_geniusroute_2019,xu_paroute_2024,xu_paroute2_2025,xu_davinci_2026,zhang_apecroute_2026}, and related models predict parasitics or generate layout constraints~\cite{ren_paragraph_2020,chen_symmetry_2021,chen_llana_2024,zhang_panda_2026,liu_refine_2026}. However, these techniques do not directly control which devices share resistive supply paths, which is a source of topology-induced mismatch. When matched devices share different wire segments carrying current from other devices, the resulting IR drops can disturb their source potentials unequally. As illustrated in Fig.~\ref{fig:info}(b), a current mirror whose reference shares a ground path with an outside device can become unbalanced, whereas placing the mirror on its own branch isolates that shared path from outside currents. The mismatch therefore depends not only on wire geometry and resistance, but also on the topology of current sharing. Geometric matching alone cannot eliminate the topology-induced mismatch, leaving a non-negligible source of post-layout performance degradation.

Routing efficiency introduces a further difficulty through the choice of routing-grid granularity. Analog routers commonly construct paths on a routing grid that is then searched by algorithms such as A*~\cite{lee_path_1961,hart_astar_1968,malavasi_area_1993}. In this context, the granularity of the routing grid is a delicate configuration. A coarse grid reduces search complexity but may not find solutions without DRC violations, while a finer grid may find the solution but increases runtime. Existing grid-shifting and refinement approaches~\cite{pan_georoute_2026} improve pin accessibility, but the problem is only partially solved by this, because reaching a pin does not necessarily produce a DRC-compliant connection. A wire may terminate with an edge misaligned with the device metal it joins, causing spacing or short-edge violations. As illustrated in Fig.~\ref{fig:info}(c), a wire reaches a pin on a coarse grid line but leaves its edge off the pin edge, whereas a wire on an edge-aligned grid line keeps its edge on the pin edge. These device-metal edges are sparse, yet a uniform grid cannot align with all of them without additional refinement. The challenge is therefore to preserve both pin accessibility and precise termination alignment without incurring the cost of a uniformly fine grid.

In this paper, we introduce \sys{}, an analog router that produces DRC-clean layouts while reconciling the three objectives through three complementary techniques. Our contributions are summarized as follows.

\begin{itemize}
  \item We propose \emph{hierarchical rule enforcement}, which assigns DRC checks and repairs to routing stages where the required geometry is complete. By covering the rules both within routing metal and where it joins device metal, it enables systematic DRC enforcement on PCell-based layouts under complex design rules from a commercial PDK.
  \item We propose \emph{topology-based mismatch mitigation} in supply tree generation, which constructs trees according to current-mirror relationships and device currents. Mirror-specific branches limit topology-induced mismatch that geometric matching cannot eliminate, while low-resistance trunks reduce voltage drops along high-current paths.
  \item We propose a \emph{hybrid routing grid strategy}, which augments a coarse routing grid with a net-specific grid aligned to device-metal edges. It preserves pin accessibility and DRC-compliant termination without the runtime cost of uniform grid refinement.
  \item Experiments on a comprehensive benchmark of \AllCircN{} circuits of different types in a TSMC 40\,nm process and a 130\,nm process show that \sys{} produces DRC-clean and LVS-correct layouts on all benchmarks under a median routing runtime of \AllRtMed{}\,s. In particular, in the TSMC 40\,nm process, \sys{} produces functional layouts with the strongest overall performance preservation, while the baselines produce no DRC-clean layouts. Ablation studies further demonstrate the contributions of the three techniques to DRC compliance, post-layout performance, and runtime.
\end{itemize}  %
\section{Preliminaries}
\label{sec:state}
\label{sec:related}

\TabCoverage
Design rules constrain the routed wires on both individual metal shapes and the geometry formed when shapes join. Rules such as spacing and via enclosure can often be checked as individual routing moves are constructed, whereas minimum-area and short-edge constraints may require the completed metal polygon. Digital routing has developed techniques for both local and polygon-level checks. TritonRoute incorporates violation markers into routing costs, while its successor uses an integrated DRC engine to check polygon rules such as minimum width and minimum step~\cite{kahng_tritonroute_2021,kahng_wxl_2022}. Dr.~CU incorporates minimum-area constraints into path search~\cite{chen_drcu_2020}, and pin-access oracles analyze feasible access points before routing~\cite{kahng_pao_2020}.

Analog routers commonly incorporate spacing constraints into their routing search or grid~\cite{chen_magicalroute_2020,zhang_sageroute_2023,pan_georoute_2026,kunal_align_2019}. Anaroute considers minimum-width, spacing, and area constraints~\cite{chen_magicalroute_2020}. GeoRoute checks wire and via moves against an R-tree of existing geometry~\cite{pan_georoute_2026}. Table~\ref{tab:coverage} compares the rule families modeled by these routers and \sys{}. Existing analog routers either leave a rule family unmodeled or model it in part, typically reducing the rule to a single value per layer and omitting the conditions it depends on, such as wire width and parallel run length for spacing or landing-metal width for via enclosure. %
\section{Algorithms}
\label{sec:algo}
\begin{figure}[tb]
  \centering
  \includegraphics{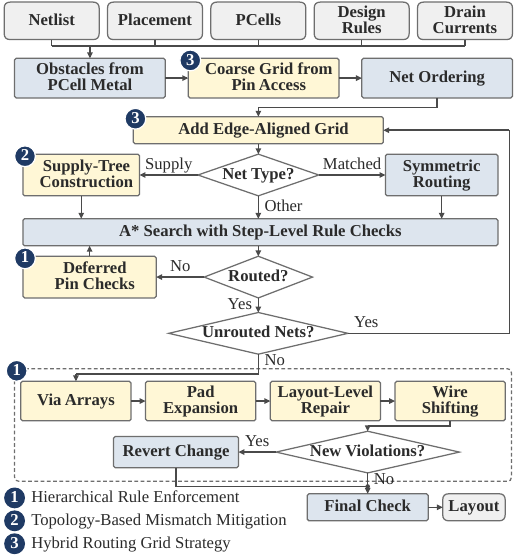}
  \caption{\CapFigFlow}
  \label{fig:flow}
  \vspace{-0.3cm}
\end{figure}

To jointly address DRC compliance, post-layout performance, and routing efficiency, \sys{} combines three key techniques, hierarchical rule enforcement, topology-based mismatch mitigation, and a hybrid routing grid strategy. Hierarchical rule enforcement distributes DRC checks and repairs across routing stages so that the rules where routing metal joins device metal are handled as systematically as those within routing metal, topology-based mismatch mitigation restricts shared supply paths between devices, and the hybrid routing grid strategy retains coarse-grid search efficiency while aligning routing wires with device-metal edges.

Fig.~\ref{fig:flow} illustrates the overall routing flow. From the netlist, placement results, PCells, design rules, and drain currents from schematic simulation, \sys{} builds obstacles from the PCell metal and sizes the coarse routing grid from pin access. Nets are then routed sequentially in a heuristic order of input-pair source connections, supply nets, matched nets, and the remaining signal nets.
For each net, \sys{} adds an edge-aligned grid to the coarse grid. Supply nets are built as supply trees, and matched nets by symmetric routing. A* search routes the supply-tree branches and the other nets, checks step-level rules, and retries a failed net with the spacing checks near its pins deferred to post-routing repair. After routing, \sys{} generates via arrays, expands pads as required by net-level rules, performs layout-level repairs, and shifts the affected wires or pads, reverting any change that introduces new violations. A final check then reports the remaining violations on the routed layout.

\subsection{Hierarchical Rule Enforcement}
\label{sec:algo-judge}
Routing metal joins device metal at pins and other routing metal at branches and vias, but the resulting geometry is not always complete during A* search. Hierarchical rule enforcement therefore assigns each design rule to a stage where the geometry needed to evaluate it is complete.
At each A* move, \sys{} checks rules on the new segment or via against surrounding metal, including width-dependent spacing and same-net spacing and notch constraints. After routing, it first repairs violations of the rules on each net's own routing metal, such as minimum-area requirements on via pads, and then those of the rules on the merged routing and device metal, including minimum-width and short-edge constraints, because the first repairs add metal that changes the merged shape. Table~\ref{tab:coverage} summarizes these rule classes and a fourth class for geometry fixed before routing.
The same checking functions evaluate A* moves and post-routing modifications. On the complete layout, they form the final check, which reports violations as records. Each post-routing modification changes only routing metal and is accepted only if it introduces no new records for the affected rules, and a violation whose repair is not accepted stays in the final check as a record. In this way, each rule is evaluated on complete geometry at its assigned stage, without requiring the A* search to anticipate geometric changes introduced by subsequent routing.

\subsubsection{Step-Level Rules}
\begin{figure}[tb]
  \centering
  \includegraphics[width=\columnwidth]{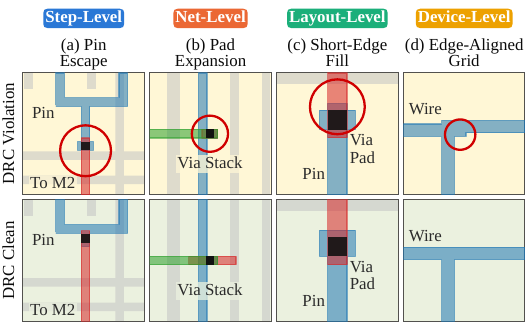}
  \caption{\CapFigMech}
  \label{fig:mech}
  \vspace{-0.3cm}
\end{figure}

Step-level rules constrain a wire segment or via relative to surrounding metal, allowing A* to check metal spacing, via enclosure, and via spacing rules at each move. Near pins embedded in dense device metal, however, strict spacing checks may prevent routing access. Refining the grid does not help, because the gap between the device metal on either side of a pin can be narrower than a wire plus the required spacing on both sides, so a wire reaching the pin violates the spacing rule at any grid position. To address this, \sys{} defers these checks to post-routing repair within a small neighborhood around each pin, in two tiers.
The first tier retains full spacing checks and attempts to escape through a via when a planar exit would violate spacing constraints. If routing fails, the second tier defers the spacing checks against other nets' pins and device metal, keeping only overlap tests during search, while retaining full checks against guard rings, stripes, other-net routing wires, and blockages. Fig.~\ref{fig:mech}(a) illustrates this partial deferral without the via escape of the first tier.  The deferred checks may leave spacing violation records near transistor terminals, which are repaired after routing by shifting the affected wires or pads slightly sideways.
Via-count requirements on wide metal are met after routing by constructing the required via arrays once the vias and their connected wide metal are complete.

\subsubsection{Net-Level Rules}
\label{sec:algo-c2}
Minimum-area rules depend on the completed metal shape of a net. After routing, \sys{} repairs their violations by expanding each violating routing-metal region, typically a bare via pad, into a containing rectangle that meets the area requirement, as shown in Fig.~\ref{fig:mech}(b). Only same-net metal is added, leaving the routed paths unchanged.

\subsubsection{Layout-Level Rules}
\label{sec:algo-c3}
Unlike net-level rules, layout-level rules depend on the geometry formed when routing metal joins device metal at pins, rings, and stripes. These include short-edge, minimum-width, and line-end rules, with line-end constraints also involving surrounding metal from other nets. Since this geometry is not fully represented by individual routing segments, the rules are checked on the complete layout and their violations are repaired through modifications to routing metal. As illustrated in Fig.~\ref{fig:mech}(c), to repair short edges, \sys{} constructs the union polygon of each net on each layer and identifies pairs of short edges sharing a corner. It then fills the rectangle bounded by the pair with same-net metal or extends the metal until one edge meets the length requirement. Similarly, \sys{} extends line ends with same-net metal to satisfy line-end constraints. Although same-net notch rules are step-level rules, notches are filled at this stage because most notches arise where routing wires merge with their own pin metal. The same sideways shifting of wires and pads follows each fill operation, where it also repairs line-end records, which exist only on the merged layout.

\subsubsection{Device-Level Rules}
Rules involving geometry fixed before routing are incorporated into the obstacles and the routing grid before A* search. All coordinates are snapped to the manufacturing grid when generated, and device metal on routing layers is treated as routing obstacles. For capacitors, routing is kept out of the capacitor's own layers and the layer above, while its marker region also prohibits repair patches. A wire terminating on device metal should also align with its edges, as shown in Fig.~\ref{fig:mech}(d).

\subsection{Topology-Based Mismatch Mitigation}
\label{sec:algo-tree}
\begin{figure}[tb]
  \centering
\providecommand{\treeK}{1.38}%
\definecolor{negbg}{HTML}{FFF7D6}\definecolor{posbg}{HTML}{EBF1DF}%
\begin{tikzpicture}[font=\footnotesize, scale=\treeK, yscale=0.75,
    dev/.style={draw, fill=black!8, minimum width=0.55cm, minimum height=0.74cm, inner sep=0pt, inner xsep=1.2pt, align=center},
    ent/.style={draw, fill=black, minimum width=0.54cm, minimum height=0.22cm, inner sep=0pt},
    w/.style={line width=1.4pt, black!45},
    wshare/.style={line width=2.0pt, red!75!black},
    wmir/.style={line width=1.4pt, blue!70!black},
    wsink/.style={line width=2.0pt, orange!85!black},
    fence/.style={dashed, blue!70!black, rounded corners=2pt},
    wtrunk/.style={line width=4.2pt, orange!85!black},
    lab/.style={font=\footnotesize, inner sep=1pt},
    tower/.pic={\draw[fill=white, line width=0.5pt] (-0.10,-0.10) rectangle (0.10,0.10);
                \foreach \x in {-0.045,0.045} \foreach \y in {-0.045,0.045} \fill (\x-0.025,\y-0.025) rectangle (\x+0.025,\y+0.025);}]
  \fill[negbg] (-0.25,-0.54) rectangle (2.90,2.80);
  \useasboundingbox (-0.25,-0.54) rectangle (6.10,2.80);%
  \fill[posbg] (2.90,-0.54) rectangle (6.10,2.80);
  \begin{scope}[shift={(0,0)}]
    \node[dev] (sa) at (0.12,2.05) {High\\Current};
    \node[dev] (ra) at (0.82,2.05) {Ref};
    \node[dev] (ma) at (2.58,2.05) {Mir};
    \node[ent] (ea) at (1.70,0.10) {};
    \node[lab, anchor=north] at (ea.south) {Root};
    \draw[w] (sa.south) -- (0.12,1.15) -- (0.82,1.15);
    \draw[w] (ra.south) -- (0.82,1.15);
    \draw[w] (ma.south) -- (2.58,1.15) -- (1.70,1.15);
    \draw[w] (1.70,1.15) -- (ea.north);
    \draw[wshare] (0.82,1.15) -- (1.70,1.15);
    \draw[-{Stealth[length=4pt]}, orange!85!black, thick] (0.90,1.30) -- (1.62,1.30);
    \node[lab, anchor=west, orange!85!black] at (1.66,1.31) {\textit{I}};
    \node[lab, anchor=north east, align=right, red!75!black] at (1.63,1.07) {Imbalanced\\IR Drop};
    \node[lab, anchor=west, red!75!black] at (0.86,1.53) {$V{+}\Delta V$};
    \node[lab, anchor=east] at (2.54,1.53) {$V$};
    \node[lab] at (1.35,-0.40) {(a) Shared Tree};
  \end{scope}
  \begin{scope}[shift={(2.95,0)}]
    \node[dev] (rb) at (0.35,2.05) {Ref};
    \node[dev] (mb) at (1.05,2.05) {Mir};
    \node[dev] (sb) at (2.35,2.05) {High\\Current};
    \node[ent, minimum width=0.84cm] (eb) at (1.50,0.10) {};
    \node[lab, anchor=north] at (eb.south) {Root};
    \draw[wmir] (rb.south) -- (0.35,1.45) -- (1.05,1.45);
    \draw[wmir] (mb.south) -- (1.05,1.45);
    \draw[wmir] (0.70,1.45) -- (0.70,0.45) -- (1.30,0.45) -- (1.30,0.18);
    \draw[wtrunk] (2.35,1.52) -- (2.35,0.45) -- (1.66,0.45) -- (1.66,0.12);
    \draw[w, orange!85!black, line width=1.4pt] (sb.south) -- (2.35,1.52);
    \pic at (2.35,1.52) {tower};
    \pic at (1.66,0.12) {tower};
    \draw[fence] (0.05,0.30) rectangle (1.40,2.47);
    \node[lab, anchor=south west, blue!70!black] at (0.05,2.47) {Mirror Branch};
    \node[lab, anchor=east, align=right, orange!85!black] at (2.24,1.00) {Trunk on\\Upper\\Layers};
    \node[lab, anchor=west, align=left] at (2.47,1.30) {Stacked\\Vias};
    \node[lab, anchor=east] at (0.31,1.28) {$V$};
    \node[lab, anchor=west] at (1.09,1.28) {$V$};
    \node[lab] at (1.35,-0.40) {(b) Supply Tree};
  \end{scope}
\end{tikzpicture}%
   \caption{\CapFigTree}
  \label{fig:tree}
  \vspace{-0.3cm}
\end{figure}

Existing geometric matching is not enough to reduce the mismatch introduced by routing. When matched devices share a supply wire carrying currents from other devices, the resulting IR drop can affect them unequally. Unlike parasitic mismatch caused by routing geometry, this topology-induced mismatch depends on which devices share resistive paths. \sys{} therefore constructs each supply net as a \emph{supply tree} that controls this sharing. Using current-mirror relationships from the netlist and drain currents from schematic simulation, both available before routing, the tree addresses two effects, imbalance within current mirrors and operating-point shifts caused by large currents.

The first effect occurs when currents from other devices flow through a current mirror's shared source path, creating unequal source voltages and disturbing its current ratio, as shown in Fig.~\ref{fig:tree}(a). To prevent this, each mirror receives a separate branch connected to the root of the supply tree, the device pin carrying the net's port label, as shown in Fig.~\ref{fig:tree}(b). The shared path within the mirror then carries only its members' currents, avoiding voltage differences caused by outside currents. To construct these branches, \sys{} identifies reference devices whose drain or source connects to their own gate net, either directly or through one other transistor channel. Devices on the supply net sharing a reference's gate net form one branch, while each remaining device forms its own. Branches are routed toward a short window of the root, and each mirror branch treats other branches as different-net metal to prevent unintended connections. Branches also treat the root pin outside the window as different-net metal, avoiding shared resistive paths along the pin strap. An ordinary routing tree subsequently connects stripes, guard rings, and branches that could not be routed independently.

While branch isolation prevents outside currents from disturbing a mirror's shared source path, it does not eliminate voltage drops caused by large currents within individual branches. To limit the resulting operating-point shifts, a branch carrying a large share of the supply current uses a low-resistance trunk, as illustrated in Fig.~\ref{fig:tree}(b). Stacked vias with full via arrays connect the device and the root to an upper routing layer, where a wire as wide as the via-array plate joins them. These trunks are reserved before routing the remaining branches. The same approach applies to output nets, where a trunk connects the two largest output-stage devices and the port pin through stacked vias.

These supply-tree constraints address IR-drop effects through routing topology and resistance, while symmetric routing further reduces the geometric mismatch between matched signal nets. \sys{} applies it to each matched pair and to each self-mirrored net with symmetric pins, of which it routes one half. The mirrored result is accepted in place of ordinary routing only if its segments pass the A* move checks and the final check reports no more violation records for any rule than the ordinary result. Supply trees take precedence over symmetry, so matched pairs containing a supply-tree net are not mirrored.

\subsection{Hybrid Routing Grid Strategy}
\label{sec:algo-grid}
\begin{figure}[tb]
  \centering
  \includegraphics[width=\columnwidth]{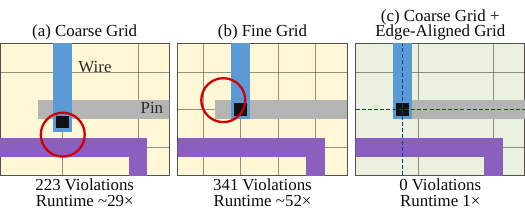}
  \caption{\CapFigGrid}
  \label{fig:grid}
  \vspace{-0.3cm}
\end{figure}
A coarse routing grid reduces search complexity but may leave wires misaligned with the device metal they connect to, causing DRC violations at their joints. A uniformly finer grid can reduce these violations, but at the cost of a larger search space. The \emph{hybrid routing grid strategy} instead retains a coarse grid and adds grid lines only where alignment is needed. Since DRC is checked on exact geometry, the grid needs to preserve only two types of locations, via sites accessible from pins and wire positions aligned with the edges of pin and power metal.

The coarse grid is first sized to preserve pin accessibility. Since layer changes are allowed only at intersections of coarse grid lines, each pin must have a reachable via site between surrounding other-net metal. The available range, or pin window, accounts for half the wire width and the required spacing on either side of a via center. Let $S$ be the largest sum of minimum wire width and spacing across the routing layers. The coarse step is then chosen as the largest multiple of $S$ that fits within the narrowest pin window, allowing the uniform grid to reach every pin.

However, a grid that provides via access does not necessarily align wires with the metal they connect to. A wire may terminate short of or beyond a device-metal edge, creating spacing violations near another net's wire, as shown in Fig.~\ref{fig:grid}(a). Misaligned terminations can also produce short edges on guard rings that cannot be filled without violating spacing constraints, as shown in Fig.~\ref{fig:mech}(d). Thus, pin accessibility alone is insufficient to ensure DRC-compliant connections.

To address these violations without uniformly refining the grid, \sys{} adds an \emph{edge-aligned grid} for each net before routing it. On each layer, an edge-aligned grid line is placed at the centerline position where a wire edge aligns with an edge of the net's own pin, guard-ring, or stripe metal, and extends across the whole routing region, as shown in Fig.~\ref{fig:grid}(c). Wires terminating on these grid lines can join device metal without creating short edges or narrow necks. Other nets' pins contribute no edge-aligned grid lines because the current net does not connect to them, and including those edges would unnecessarily densify the grid. Alignment therefore no longer constrains the step, which is set by pin accessibility alone, and each net adds only one grid line per edge of its own pin and power metal, whereas a uniform refinement by a factor of $k$ multiplies the grid nodes of every net by $k^2$. The resulting net-specific grid preserves the coarse search space while providing the alignment needed at routing terminals.
\section{Experimental Results}
\label{sec:exp}
\TabBOne

\subsection{Experimental Setup}
\label{sec:exp-setup}

\subsubsection{Benchmarks and evaluation}
We evaluate \sys{} on a comprehensive benchmark with 11 circuits in a TSMC 40\,nm process and \OthSkyN{} circuits in the SkyWater 130\,nm process~\cite{skywater_pdk_2026}, including OTAs, comparators, low-dropout regulator (LDO), and voltage-controlled oscillator (VCO). We used an open-source analytical placer APPlace~\cite{liu_applace_2027} to produce the placement. Layouts are verified using Calibre DRC and LVS under the design rules of the TSMC 40\,nm PDK, and using Magic and KLayout DRC and Netgen LVS for SkyWater 130\,nm. We report routing-side DRC violations, defined as violations introduced by routing relative to the placed layout, excluding chip-level rules, including seal rings, required marker layers, metal density, and ESD checking.
We then use Calibre to extract the parasitics and Spectre to evaluate each layout at the typical process corner using a common testbench. The measured metrics are phase margin (PM), unity-gain bandwidth (UGB), DC gain, common-mode rejection ratio (CMRR), and input-referred offset, and for the comparator, delay, power, and input-referred offset. The CMRR and offset of the schematic are ideal, so they are omitted in Table~\ref{tab:res}. A layout is considered functional if it is LVS-correct and remains stable after extraction, with valid PM, UGB, and positive loop gain at 1\,Hz under closed-loop operation at the amplifier's gain. The target PM is at least 45 degrees. Pre-layout performance from schematic simulation is shown for reference.

\subsubsection{Baselines}
We compare \sys{} on the same APPlace placements with two existing analog routers, Anaroute~\cite{chen_magicalroute_2020,anaroute_github_2026}, the open-source router of MAGICAL, and SAGERoute~\cite{zhang_sageroute_2023}, run from the binary. Anaroute uses MAGICAL-generated devices positioned at the device centers of APPlace, since it is designed only for those self-generated devices. PaRoute~\cite{xu_paroute_2024} and PaRoute2~\cite{xu_paroute2_2025} employ learned guidance on top of Anaroute's routing kernel, which is another orthogonal problem, so we used Anaroute as the relevant baseline for comparing routing mechanisms. ALIGN~\cite{kunal_align_2019,dhar_align_2021,align_github_2026} relies on grid-aligned device generation, which differs from the PCell-based routing setting evaluated here and it cannot deliver LVS clean layout with PCells from commercial PDKs, and is therefore not included in the comparison. 

\TabComp
\BlockTwo

\subsection{Router Comparison}
\label{sec:exp-drc}
As shown in Fig.~\ref{fig:stats}, \sys{} produces DRC-clean, LVS-correct, and functional layouts for all 11 circuits in TSMC 40\,nm with the shortest median routing runtime, whereas neither baseline produces a DRC-clean layout and each produces functional layouts on at most three circuits. Table~\ref{tab:res} details the results on the ten OTAs in DRC compliance, post-layout performance, and runtime. On the comparator, as shown in Table~\ref{tab:comp}, only \sys{} produces a DRC-clean layout, with the shortest post-layout delay.
\FigStats

\subsubsection{DRC Compliance}
\label{sec:exp-b-drc}
As shown in Table~\ref{tab:res}, \sys{} produces LVS-correct layouts with no routing-side DRC violations on all ten OTAs. In contrast, the only two LVS-correct layouts of Anaroute contain \SelRouterLVSDRCMin{} and \SelRouterLVSDRCMax{} violations, and its LVS-incorrect layouts contain violations of the same order of magnitude. SAGERoute leaves \SelSageRoutMin{} to \SelSageRoutMax{} violations and fails LVS on every circuit it completes. The DRC-clean results of \sys{} follow from hierarchical rule enforcement, which enforces each rule at the stage where its geometry is complete. It thereby covers the rules within routing metal as well as those where routing metal joins device metal, which spacing checks during search alone do not cover.

\subsubsection{Post-Layout Performance}
\label{sec:exp-b-perf}
\sys{} produces functional post-layout circuits on all ten OTAs, compared with \SelRouterUsableMagP{} for Anaroute and none for SAGERoute. All layouts of \sys{} remain stable with a phase margin of at least 45 degrees, while their DC gains deviate from the schematic by at most \SelTreeDCchgAbsMax{}\,dB. These results show that \sys{} preserves the circuit behavior established during sizing better than the baselines.
They are also consistent with the two goals of topology-based mismatch mitigation, preventing unrelated currents from disturbing the shared source paths of current mirrors and reducing voltage drops along high-current paths through low-resistance trunks.

On ota1 and ota8, where both \sys{} and Anaroute produce functional layouts, \sys{} achieves higher phase margins and DC gains, while each router achieves a higher UGB on one circuit. Anaroute achieves higher CMRR and lower input-referred offset on both. Across the full benchmark, \sys{} provides the only functional layouts on \SelRouterOnlyUsable{} circuits and maintains comparable post-layout performance on the remaining two. This gives \sys{} the strongest overall performance preservation among the evaluated routers.

\subsubsection{Runtime}
\label{sec:exp-b-time}
As shown in Table~\ref{tab:res}, \sys{} achieves a median routing runtime of \SelRtTreeMed{}\,s, compared with \SelRtMagPMed{}\,s for Anaroute and \SelRtSageMed{}\,s for SAGERoute. Although Anaroute is faster on \SelRtFasterMagPN{} circuits, those layouts contain at least \SelRtFasterMagPDRCMin{} DRC violations each, and only ota8 is LVS-correct. The runtime efficiency of \sys{} comes from its coarse routing grid, whose step is set to $2S$, the largest multiple of $S$ allowed by the narrowest pin window across the benchmarks. As described in Section~\ref{sec:algo-grid}, a net-specific edge-aligned grid provides the required alignment with device-metal edges without uniformly refining this grid, keeping the routing search space small while maintaining DRC compliance.

\subsection{Ablation Study}
\label{sec:exp-abl}
Each row of Table~\ref{tab:abl} disables one mechanism while keeping the remaining routing flow unchanged, and the rows are grouped by the three techniques of Section~\ref{sec:algo}. The rule-enforcement rows test whether each of the step-level, net-level, and layout-level stages is needed for DRC compliance, the mismatch-mitigation rows identify what the supply tree and its trunks contribute to post-layout performance, and the grid rows test whether refining the uniform grid can replace the edge-aligned grid in DRC compliance and runtime. To keep the post-layout metrics consistent, the ablations are reported on the ten OTAs.

\subsubsection{Hierarchical Rule Enforcement}
\label{sec:exp-abl-rule}
As shown in Table~\ref{tab:abl}, removing any one of the rule-enforcement mechanisms leaves \AblStrictDRC{} to \AblPadDRC{} routing-side DRC violations, compared with \AblBaseDRC{} in the full configuration. The largest increase, \AblPadDRC{} minimum-area violations, occurs without pad expansion, because step-level spacing checks cannot enforce area constraints on completed routing shapes. Deferring all checks at pin escape also introduces shorts and an LVS failure, showing that the deferral needed for pin access has to stay limited to pins and device metal. Other violations arise from missing via arrays and layout-level repairs, even when step-level checks remain active. These results show that search-time checks alone cannot replace the later stages, since each stage addresses geometric constraints that the others do not cover.

\subsubsection{Topology-Based Mismatch Mitigation}
\label{sec:exp-abl-tree}
\begin{figure}[tb]
  \centering
  \includegraphics[width=\columnwidth]{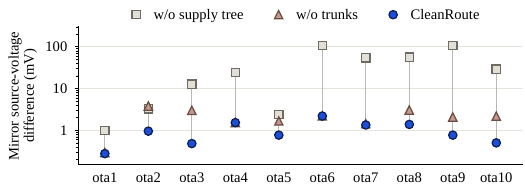}
  \caption{\CapFigIrdrop}
  \label{fig:irdrop}
  \vspace{-0.3cm}
\end{figure}
Removing the supply tree reduces the number of stable layouts from \SelAmps{} to \AblPostTreeStable{}, with only \AblPostTreePMtarget{} retaining a PM of at least 45 degrees. On the \AblPostTreePairN{} circuits stable in both configurations, the median input-referred offset increases from \AblPostTreePairBaseOffMedPr{} to \AblPostTreePairOffMed{}\,\textmu V. These changes are consistent with the source-voltage imbalance introduced when currents from unrelated devices share the resistive paths of current mirrors. The post-layout DC operating point shows this imbalance. As shown in Fig.~\ref{fig:irdrop}, the largest source-voltage difference between a mirror's reference and its copies on the ground net has a median of \IrTreeMed{}\,mV without the supply tree and \IrBaseMed{}\,mV with it, and it is lower with the supply tree on \IrLowerN{} circuits. Since symmetric routing remains available in this ablation, the degradation demonstrates that geometric matching alone is insufficient to preserve the same post-layout behavior.

Low-resistance trunks address a complementary source of IR drop along paths carrying large currents. Without them, all \SelAmps{} circuits remain stable with PMs of at least 45 degrees, but the number retaining a PM of at least 60 degrees falls from \AblPostBasePMsixty{} to \AblPostTrunkPMsixty{}. Together, the two ablations distinguish the roles of mirror-aware branches and heavy-current trunks, where the former limit current-sharing imbalance and the latter reduce resistive drops that can degrade circuit performance.

\subsubsection{Hybrid Routing Grid Strategy}
\label{sec:exp-abl-grid}
Removing the edge-aligned grid while retaining the coarse grid leaves \AblFlushDRC{} routing-side violations, with \AblFlushLVS{} of the \SelCircuits{} layouts passing LVS and the median runtime rising from \AblBaseTime{} to \AblFlushTime{}\,s, as illustrated in Fig.~\ref{fig:grid}(a). The runtime rises because connections that fail the checks at misaligned positions are retried many times. The grid still provides pin access, but its wires need not align with the edges of the device metal they join. This separates two requirements that a uniform grid treats together, reaching a pin and forming a DRC-compliant connection to it.

Refining the grid does not remove this distinction. Reducing the step from $2S$ to $S$ while retaining the edge-aligned grid increases the median runtime from \AblBaseTime{} to \AblGridSTime{}\,s and still leaves \AblGridSDRC{} violations. Without the edge-aligned grid, even this finer grid leaves \AblNoFlushSDRC{} violations, mostly short edges, with a median runtime of \AblNoFlushSTime{}\,s, as illustrated in Fig.~\ref{fig:grid}(b). Uniform refinement is therefore neither necessary nor sufficient for DRC compliance in these experiments. The edge-aligned grid introduces the alignment coordinates needed at each net's connections, allowing the coarse grid to achieve zero routing-side violations without the runtime cost of uniform refinement.

\subsubsection{Different Placers}
\label{sec:exp-abl-place}
The preceding experiments use APPlace~\cite{liu_applace_2027} as the placer. To examine whether the results depend on this placement methodology, we also route the ten circuits placed by MAGICAL's complete open-source flow~\cite{xu_magical_2019,zhu_tutorial_2023,magical_github_2026}. With the same routing configuration, \sys{} produces functional layouts for all \SelMPFunc{} circuits and leaves no routing-side DRC violations on \SelMPCleanN{} of them, as shown in Table~\ref{tab:swap}. The remaining \SelMPDirtyDRC{} violations on \SelMPDirtyName{} occur at one location, where an output wire and a ground wire violate third-layer spacing, and are also detected by the final check. The results extend the observed routing reliability beyond the placement methodology used in the main comparison.

\subsection{PDK Generalization}
\label{sec:exp-other}
\TabOther
We further run \sys{} on the \OthSkyN{} circuits in SkyWater 130\,nm, with no algorithm adapted and only the rule values in the PDK replaced. As shown in Table~\ref{tab:other}, all \OthSkyN{} layouts are DRC-clean and LVS-correct, with a median routing runtime of \OthSkyRtMed{}\,s. The baselines cannot be run on SkyWater 130\,nm because their source code hard-codes the layer stack and device structures, so no comparison is available, and post-layout simulation results are therefore omitted. This result shows that hierarchical rule enforcement handles design rules systematically. Each rule is modeled at the stage where its geometry is complete, with its values taken from the PDK, so a new process requires no change to the routing flow.
\section{Conclusion}
\label{sec:conclusion}

In this paper, we presented \sys{}, an analog router that reconciles DRC compliance, post-layout performance preservation, and routing efficiency. Experiments on a comprehensive benchmark of \AllCircN{} circuits of different types in a TSMC 40\,nm process and a 130\,nm process show that \sys{} produces DRC-clean and LVS-correct layouts on all benchmarks with a median routing runtime of \AllRtMed{}\,s. In the TSMC 40\,nm process, \sys{} produces functional layouts with the strongest overall performance preservation. 
\ifdefined\arxiv\else\clearpage\fi%
\bibliographystyle{IEEEtran}
\bibliography{./ref/router}

\end{document}